\documentclass[11pt,oneside]{article}
\usepackage{setspace,graphicx,amssymb,amsmath,latexsym,amsfonts,amscd,amsthm,multirow,ctable,mathdots,caption,array}
\usepackage{fancyhdr,tabularx,cite,mathrsfs}
\usepackage[headings]{fullpage}
\usepackage{stmaryrd}
\usepackage{rotating}
\usepackage{authblk}
\usepackage{hyperref}

\newcommand{\bea}{\begin{eqnarray}} 
\newcommand{\eea}{\end{eqnarray}} 
\newcommand{\bee}{\begin{eqnarray*}} 
\newcommand{\eee}{\end{eqnarray*}} 
\newcommand{\al}{\begin{align*}} 
\newcommand{\eal}{\end{align*}} 
\newcommand{\be}{\begin{equation}} 
\newcommand{\ee}{\end{equation}} 
 
\newcommand{\bem}{\begin{pmatrix}} 
\newcommand{\eem}{\end{pmatrix}}

\newcommand{\cN}{{\mathcal N}}

\def\a{\alpha}

\def\t{\tau} 
\def\th{\theta} 
\def\til{\tilde}

\newcolumntype{R}{ >{$}r <{$}}
\newcolumntype{C}{ >{$}c <{$}}
\newcolumntype{L}{ >{$}l <{$}}
\newcolumntype{F}{>{\centering\arraybackslash}m{1.5cm}}

\def\ll{\ell}

\newcommand{\comment}[1]{}

\newcommand{\tr}{\operatorname{{tr}}}

\newcommand{\SU}{\operatorname{\textsl{SU}}}    

\theoremstyle{definition}

\theoremstyle{remark}

\numberwithin{equation}{section}

\begin{document}

\setstretch{1.4}

\title{
\vspace{-35pt}
    \textsc{\huge{Extremal chiral $\mathcal N=4$ SCFT with $c=24$
    }  }
}
\author{Sarah M. Harrison\thanks{sarharr@physics.harvard.edu}}

\affil{Center for the Fundamental Laws of Nature\\ 
Harvard University, Cambridge, MA 02138, USA}
\date{}

\maketitle

\abstract{
We construct an extremal chiral $\mathcal N=4$ superconformal field theory with central charge 24 from a $\mathbb Z_2$ orbifold of the chiral bosonic theory with target $\mathbb R^{24}/\Lambda$, where $\Lambda$ is the Niemeier lattice with root system $A_2^{12}$. This construction is analogous to constructions of extremal chiral $\mathcal N=1$ and $\mathcal N=2$ CFTs with $c=24$, where $\Lambda = \Lambda_{Leech}$ and the Niemeier lattice with root system $A_1^{24}$, respectively. The theory has a discrete symmetry group related to the sporadic group $M_{11}$.
}

\clearpage

\tableofcontents

\clearpage

\section{Introduction}
One of the most outstanding problems in physics today is understanding the nature of quantum gravity. The AdS/CFT correspondence has led to many insights in recent years, as it has furnished a number of exact dualities between quantum theories of gravity in negatively curved space and conformal field theories in one lower dimension. Many of these examples arise as solutions of string theory and M-theory\cite{Magoo}. It has also inspired progress in the understanding of black holes and the quantum gravity S-matrix, among other things.

The context of an extremal CFT was originally introduced by Witten in \cite{Witten} (see also \cite{hoehn}) as an attempt to understand potential holographic duals to pure (super)gravity in $AdS_3$. In a sense, this is the simplest possible example of a holographic duality. The idea is, with a few assumptions, to use  modular properties of the 2d CFT and the holographic principle to fix the full quantum gravity partition function in anti-de Sitter space. Because there are no metric degrees of freedom in three dimensions, he assumes that solutions which contribute to the quantum gravity path integral are the vacuum and BTZ black holes, and that the latter contribute to the CFT partition function as primaries with dimension greater than $c/24$; with this information, and the fact that the leading term in the expansion of the partition function around $\tau \to i \infty$ is $q^{-c/24}$, $SL(2,\mathbb Z)$ invariance fixes the function to be of the form
\be
Z_k(\t) = q^{-k}\prod_{n=2}^\infty {1\over 1-q^n} + \mathcal O(q),
\ee
where $k\geq 1$ is an integer parametrizing the allowed values of the central charge, $c=24k$, and the infinite product comes from the Virasoro descendants of the vacuum.

Witten also makes a similar prediction for the partition functions of candidate duals to pure supergravity in $AdS_3$ in terms of an integer $k^*\geq 1$ which parametrizes the allowed values of the central charge, now $c=12k^*$. In this case, the only difference is that now the partition function has slightly modified modular properties which take into account the spin structure of the fermionic states on the torus, and the descendants of the vacuum are generated by the full $\mathcal N=1$ enhancement of the Virasoro algebra. Nevertheless, this is enough to fix the partition functions $Z_{k^*}(\t)$, as it was in the bosonic case.

One immediate surprise is that there are known chiral conformal theories with these partition functions for $k=1$ and $k^*=1,2$, and that they have interesting connections to the theory of sporadic groups, unimodular lattices, and error correcting codes. The theory for $k=1$ is the famous monster CFT of Frenkel, Lepowski, and Meurman \cite{FLM}, which is a theory constructed from a $\mathbb Z_2$ orbifold of chiral bosons on the Leech lattice, and whose symmetry group is the monster group, the largest of the sporadic finite simple groups. The theories for $k^*=1$ and $2$ are also interesting chiral theories which can be realized as lattice orbifolds and with large symmetry groups related to the Conway group $Co_0$,  the automorphism group of the Leech lattice. The theory with $k^*=1$ was originally constructed in \cite{FLM} as a $\mathbb Z_2$ orbifold of eight bosons compactified on the $E_8$ root lattice with their fermionic superpartners; properties of this theory were further studied in \cite{Duncan,DM-C}. A theory with $k^*=2$ was  constructed by Dixon, Ginsparg, and Harvey \cite{DGH} as a nonlocal $\mathbb Z_2$ orbifold of the Leech lattice; this construction will be described in more detail in \S \ref{sec:DGH}.

One of the crucial assumptions underlying the predictions in \cite{Witten} is the holomorphic factorization of the  partition functions of the dual CFTs. In \cite{MalWit}, the authors studied what happens when one relaxes this assumption from the perspective of the gravitational path integral; this leads to some confusion, one possible resolution of which could be that these theories do not exist in general for large $k$. Other investigations into extremal theories with $k>1$ have been similarly inconclusive, in that their existence has neither been proven nor ruled out \cite{yin1,yin2,gaiottoyin,gab,gabetal,gabkel,gaiotto}. Finally, it has been suggested that one may interpret chiral conformal field theories as duals to a theory of ``chiral gravity" in $AdS_3$ \cite{LSS}; the viability of these theories is somewhat controversial, but that discussion is not relevant to this paper.

The concept of an extremal CFT was extended to theories with $\mathcal N=2$ and $\mathcal N=4$ superconformal symmetry in \cite{GGKMO}. The extremal condition imposes a  constraint on the elliptic genus, which, for a 2d $\mathcal N=(2,2)$ theory, is given by
\be
Z_{EG}(\t,z) = \tr_{\rm RR} (-1)^{F} e^{2\pi i z J_0} q^{L_0 - c/24} \bar q^{\bar L_0 - \bar c/24},
\ee 
where $J_0$ is the left-moving $U(1)$ charge, $c$ and $\bar c$ are the left- and right-moving central charges, respectively, and the trace is taken in the Ramond-Ramond sector of the theory. $Z_{EG}$ is a purely holomorphic function of $\tau$ when the theory has a compact target space and is given by a weak Jacobi form of weight zero and index $c/6$\cite{KYY}. It can be expanded as
\be
Z_{EG}(\t,z) = \sum_{n\geq 0,\ell} c(n,\ell) q^n y^\ell
\ee
where $y = e^{2\pi i z}$. The extremality condition is a constraint on the coefficients $c(n,\ell)$ of terms in the elliptic genus with negative ``polarity" $p$,  that is, states for which $p=4mn -\ell^2 <0$. Here $m=c/6$ is the index of the Jacobi form. States with negative polarity can be viewed as states which violate the cosmic censorship bound \cite{Farey}; thus, if one assumes the only such states arise from the NS vacuum and its $\mathcal N=2$ (or $\mathcal N=4$) descendants, one has a constraint on all of the polar coefficients in the elliptic genus. 

It turns out that this constraint is usually more restrictive for a given $m$ than the space of Jacobi forms at that index allows, and there exists only a handful of extremal $\mathcal N=2$ and $\mathcal N=4$ elliptic genera under this definition\cite{GGKMO}. It is an interesting question to consider modifying this condition to some notion of ``near extremality," where a bounded number of states contributes to the index up to a  dimension parametrically smaller than $c$ as the central charge gets large. This is considered somewhat in \cite{GGKMO}, but it still merits further investigation.

One advantage of the notion of extremality for theories with extended supersymmetry is that, unlike in the case of $\mathcal N=0$ and $\mathcal N=1$ theories, one does not have to assume holomorphic factorization of the 2d CFT. However, as already mentioned, there is only a finite number of candidate elliptic genera satisfying this condition with very small central charge. Additionally, the only known constructions of theories satisfying these extremality conditions happen to be chiral in any case, and have partition function equal to the would-be extremal elliptic genus at the corresponding left central charge. For central charge 12 ($m=2$), the chiral CFTs discussed in \cite{M5} furnish examples of extremal $\mathcal N=2$ and $\mathcal N=4$ theories, and, similarly, the chiral theories constructed in \cite{extN=2} and this paper are examples of extremal $\mathcal N=2$ and $\mathcal N=4$ theories with central charge 24 ($m=4$), respectively. These theories, along with the $k=1$ and $k^*=1,2$ theories introduced in \cite{Witten} are the only known extremal CFTs.\footnote{Note that the $c=12$ theory first discussed in \cite{spin7} and analyzed in detail in \cite{exceptional} satisfies an analogous type of extremal constraint, in that there are no (NS) primaries above the vacuum of dimension $\leq c/24$. This theory has an extended chiral superconformal algebra, known sometimes as the $\mathcal{SW}(3/2,2)$ algebra, or the algebra associated to string compactification on manifolds of Spin(7)holonomy. It could be of interest to analyze extremal constraints for theories with extended chiral algebras such as $W$-algebras or super-$W$-algebras. This was in fact originally mentioned in \cite{Witten}.}\footnote{It is not clear to me whether one should count the $K3$ conformal field theory as an extremal CFT, as, though it is proportional to the $m=1$ extremal elliptic genus of \cite{GGKMO}, it seemingly does not have the right overall coefficient as predicted in that paper, at least for the case of $\mathcal N=2$.} It would be very interesting to find the first example of such a theory with $c>24$ with any amount of supersymmetry.

The rest of the paper is organized as follows. In \S \ref{sec:EG} we discuss the definition of the extremal elliptic genus in more detail, reviewing the work of \cite{GGKMO} and focusing on the case of the extremal $\mathcal N=4$ genus for $m=4$. In \S \ref{sec:DGH} we review the salient points of \cite{DGH} necessary for constructing a chiral superconformal field theory from a $\mathbb Z_2$ orbifold of a unimodular lattice. We explicitly construct the extremal chiral $\mathcal N=4$ theory with $c=24$ in \S \ref{sec:theory} by adapting the methods of \cite{DGH} to an orbifold of chiral bosons on the Niemeier lattice with root system $A_2^{12}$. That is, we show that from this orbifold theory, one can construct a $c=24$ $\mathcal N=4$ superconformal algebra with the correct OPEs, and that the chiral partition function of this theory is precisely the extremal elliptic genus with $m=4$. Finally we conclude in \S \ref{conclusion} with a number of interesting questions raised given the recent growth in the collection of extant extremal CFTs. A number of appendices collect useful formulae, character decompositions of the first few proposed extremal elliptic genera, further details on the ternary Golay code and its connection to the Mathieu group $M_{12}$, and some explicit details of calculations involved in the construction of the $\mathcal N=4$ superconformal algebra.

\section{Extremal elliptic genera}\label{sec:EG}
Recall that a weak Jacobi form of weight $k$ and index $m$ is a function $\phi_{k,m}(\tau,z)$ on $\mathbb H\times \mathbb C$ with the following modular transformation property,
\be
\phi_{k,m} \left ({a \tau + b\over c\tau +d}, {z \over c\tau + d}\right ) = (c\tau + d)^k e^{2 \pi i m {c z^2\over c\tau + d}} \phi_{k,m}(\tau, z)~~~ \forall \left(\begin{array}{cc}
a & b  \\
c & d  \end{array}\right) \in SL(2, \mathbb Z),
\ee
and elliptic transformation property
\be
\phi_{k,m}(\tau, z + \ell \tau + \ell') = e^{-2 \pi i m(\ell^2\tau + 2\ell z)} \phi_{k,m}(\tau,z)~ ~~ \forall \ell, \ell' \in \mathbb Z.
\ee
The ring of weak Jacobi forms of weight zero and index $m$ has four generators: the two Eisenstein series $E_4(\tau)$ and $E_6(\tau)$, and two weak Jacobi forms, $\varphi_{-2,1}(\tau,z)$ and $\varphi_{0,1}(\tau,z)$, of weight and index, $(k,m)$, equal to $(-2,1)$ and $(0,1)$, respectively. Formulas for these functions are given in appendix \ref{app:formulas}. For more information about the properties of Jacobi forms, see, e.g., \cite{EZ}.

In appendix \ref{app:functions} we list the first several cases of extremal $\mathcal N=2$ and $\mathcal N=4$ Jacobi forms, as first discussed in \cite{GGKMO}. We also give their decompositions into superconformal characters, neglecting an overall constant coming from the degeneracy of right-moving Ramond ground states which is derived in \cite{GGKMO}. The extremal $\mathcal N=2$ and $\mathcal N=4$ elliptic genera coincide for $m=1,2$. For $m=3,4,5,$ they differ, and for $m >5,$ the extremal $\mathcal N=4$ elliptic genus very likely doesn't exist \cite{GGKMO}. 

Note that for the cases of even $m$, the extremal $\mathcal N=2 (4)$ elliptic genus has the potential to be interpreted as a partition function of a chiral $\mathcal N=2(4)$ conformal field theory, since the coefficients in the character expansion are all positive. In this case, the constant coming from the degeneracy of right-moving Ramond ground states normally present in the elliptic genus must be one. In fact, for the case of $m =2,4$,  corresponding extremal theories have been constructed as chiral CFTs with $\mathcal N=2$ or $\mathcal N=4$ superconformal symmtery. This paper provides a construction of an $\mathcal N=4$ ECFT with $m=4$; see \cite{M5} for the case of the $\mathcal N=2,4$ ECFTs for $m=2$ and \cite{extN=2} for the $\mathcal N=2$ theory with $m=4$.  

However, this property does not hold for the case of odd $m$. As one can see from the decompositions in the appendix, there is always a negative contribution coming from a short multiplet in the Ramond sector. In the case of the $\mathcal N=2$ extremal elliptic genus, the multiplet with a negative coefficient is a BPS primary of dimension and $U(1)$ charge $h= {m\over 4}$ and $Q=m$; in the $\mathcal N=4$ case, it is a BPS primary of dimension and ${1\over 2}J_3$ charge $h={m\over 4}$ and $j={m\over 2}$. One can check that this property also holds for the proposed extremal $\mathcal N=2$ elliptic genera of $m= 7,8,11,13$ which are not given in the appendix. Thus for cases of odd $m$, if an extremal theory exists, it will necessary not be chiral.

We will consider the extremal $\mathcal N=4$ elliptic genus for the case of $m=4$. The relevant weight zero Jacobi form is
\be
Z^{m=4}_{\mathcal N=4}(\tau,z)= {271 \over 576}E_4^2 \varphi_{-2,1}^4 + {43\over108}\varphi_{0,1}\varphi_{-2,1}^3 E_6 + {37\over 288} \varphi_{0,1}^2\varphi_{-2,1}^2E_4+ {5 \over 1728}\varphi_{0,1}^4.
\ee
The decomposition into $c=24$, $\mathcal N=4$ superconformal characters can be written
\bea\nonumber
Z^{\mathcal N=4}_{m=4}(\tau,z) &=&55 {\rm ch}_{{5}; {1},0}(\t,z) + {\rm ch}_{{5}; {1},2}(\t,z)\\\nonumber
&+& (18876 + 1315512 q + \ldots)( {\rm ch}_{{5}; {2},{1\over 2}}(\t,z)+{\rm ch}_{{5}; {2},-{1\over 2}}(\t,z))\\\nonumber
&+& (12045 + 1152943 q + \ldots)( {\rm ch}_{{5}; {2},1}(\t,z)+{\rm ch}_{{5}; {2},-1}(\t,z))\\\nonumber
&+& (1980 + 391974 q + \ldots)( {\rm ch}_{{5}; {2},{3\over 2}}(\t,z)+{\rm ch}_{{5}; {2},-{3\over 2}}(\t,z))\\\label{eq:N=4exp}
&+& (33 + 45990 q +  \ldots) {\rm ch}_{{5}; {2},2}(\t,z),
\eea
where ${\rm ch}_{{m+1}; {h},j}(\t,z)$ is the $\mathcal N=4$ superconformal charater of central charge $6m$ in the Ramond sector for a primary of dimension $h$ and ${1\over 2}J_3$ charge $j$. The primaries with $h=1$ are BPS (short) multiplets and those with $h >1$ are non-BPS (long) multiplets. Note that for the long multiplets, the following character identity holds:
\be
{\rm ch}_{{5}; {2+n},{j}}(\t,z)=q^n{\rm ch}_{{5}; {2},{j}}(\t,z), ~~\forall n\in \mathbb Z,
\ee
so we  write the character expansion for the long multiplets entirely in terms the characters for the states with dimension $h=2$ times a power series in $q$. See appendix \ref{app:formulas} for the structure of $\mathcal N=2$ and $\mathcal N=4$ superconformal characters.

On the face of it, the coefficients in equation (\ref{eq:N=4exp}) may not seem particular meaningful; however, in the course of constructing a chiral CFT with this partition function, we will prove that they are  naturally related to dimensions of irreducible representations of the sporadic Mathieu group $M_{11}$.

\section{Chiral SCFTs from orbifolds of self-dual lattices}\label{sec:DGH}
In this section we review the work of Dixon, Ginsparg, and Harvey \cite{DGH}, in which they show how to construct a $c=24$ chiral conformal field theory with $\mathcal N=1$ supersymmetry from a $\mathbb Z_2$ orbifold of an even, unimodular lattice of dimension 24.

Bosonic (chiral) conformal field theories with modular invariant partition functions can be constructed from even, self-dual, unimodular lattices with rank $24k$, for $k$ an integer. In this case the CFT will have central charge $24k$. In dimension 24, there are 24 such lattices, the Leech lattice, the unique even, self-dual, unimodular lattice with no roots (vectors of length-squared 2), and the 23 Niemeier lattices, which can be classified by their root systems, which are a union of simply-laced root systems of the same Coxeter number. The authors of \cite{DGH} showed how one can construct a chiral $\mathcal N=1$ superconformal field theory from one of these lattices by considering a $\mathbb Z_2$ orbifold of the theory.

For the moment we leave the choice of lattice arbitrary. The partition function of the un-orbifolded chiral theory is given by
\be
Z_{\Lambda}(\tau)= {\Theta_\Lambda(\tau)\over \eta(\tau)^{24}} = J(\tau) + (c_\Lambda+24) = {1\over q} + (c_\Lambda+24) + 196884 q + \ldots,
\ee
where $\Theta_\Lambda$ is the lattice theta function, $c_\Lambda$ is the number of roots of the lattice, and $J(\tau)$ is the unique weight zero modular invariant function such that $J(\tau)\sim {1\over q} + O(q)$ as $\tau \to i\infty$. The primary fields of the theory include the vacuum and products of 24 dimension one currents $J_i= i\partial x_i$ for $i=1,...24$, and vertex operators of dimension $\beta^2/2$ of the form
\be
V_\beta(\tau) = :e^{i\beta\cdot x(\tau)}:,~~ \forall \beta \in \Lambda.
\ee
There also exists a dimension two stress tensor given by
\be
T(z)=-{1\over 2}:\partial x_i\partial x_i:(z)
\ee
whose modes generate a Virasoro algebra.

We want to consider the theory under a $\mathbb Z_2$ orbifold which acts as $g: x_i \to -x_i ~\forall i$. The untwisted sector Hilbert space $\mathcal H$ splits into two spaces which we define as
\be
\mathcal H_\pm := \{\psi \in \mathcal H | g\psi = \pm \psi\},
\ee
based on whether the state is invariant or anti-invariant under the orbifold action. There is also a twisted Hilbert space $\mathcal H^{tw}$ introduced by the action of the orbifold, which can also be decomposed in a similar way:
\be
\mathcal H^{tw}_\pm := \{\psi^{tw} \in \mathcal H^{tw} | g\psi^{tw} = \pm \psi^{tw}\}.
\ee

The fields in $\mathcal H^{tw}$ satisfy boundary conditions $x^i(e^{2\pi i}z)= - x^i(z)$. There are $2^{12}$ holomorphic twist fields $\sigma^a$ which arise from the $2^{24}$ fixed points $\mathbb Z_2$ orbifold, where the holomorphic projection essentially acts as a square root on the number of fixed points. The action of these twist fields on the vacuum state in the untwisted sector produces $2^{12}$ degenerate twisted sector ground states with  dimension $h=3/2$. That is, they have OPE with the stress tensor,
\be
T(z) \sigma^a(w) \sim {3\over 2}{\sigma^a(w)\over (z-w)^2} + \ldots
\ee
and OPEs with each other which go like
\be
\sigma^a(z) \sigma^b(w) \sim {\delta^{ab}\over (z-w)^3} + \ldots.
\ee
The twisted sector Hilbert space $\mathcal H^{tw}$ is composed of the $\sigma^a$ acting on the untwisted primaries, together with their descendants coming from the action of twisted oscillator modes.

The Hilbert space of the bosonic twisted theory is given by projecting onto $\mathbb Z_2$ invariant states which lie in the untwisted and twisted Hilbert spaces: $\mathcal H_+ + \mathcal H^{tw.}_+$. The projection picks out the twisted sector states which have mutally local OPEs. These twisted theories are discussed in, e.g., \cite{schell,mont}. The partition function is given by
\be
Z_\Lambda^{tw}(\tau) = J(\tau) + 12h_\Lambda,
\ee
where $h_\Lambda$ is the coxeter number of the root system of the lattice $\Lambda$, with $h_\Lambda=0$ when $\Lambda= \Lambda_{Leech}$. Notice that for the case of the Leech lattice theory, the orbifold projects out the 24 dimension 1 currents $i\partial x_i$, and the resulting theory is rigid and has an automorphism group which is the monster group\cite{FLM}.

The authors of \cite{DGH} showed that one can instead construct an $\mathcal N=1$ superconformal theory by considering the full Hilbert space of the orbifolded theory including both invariant and anti-invariant states. The first step is to note that the fields $\sigma^a$ have the correct dimension for a supercurrent. One then splits the Hilbert space into two sectors based on whether the OPE with the supercurrent has a branch cut (R sector) or not (NS sector.) One then finds the Hilbert space of the NS sector is
\be
\mathcal H_{NS} = \mathcal H_+ + \mathcal H^{tw.}_-,
\ee
and the R sector is
\be
\mathcal H_{R} = \mathcal H_- + \mathcal H^{tw.}_+.
\ee

In the case of $\Lambda$ the Leech lattice, there are no primary fields of dimension one in the NS sector, so there is no way to construct a current algebra needed for extended supersymmetry. However, for any of the other Niemeier lattices, there are dimension one primary fields which one can use to construct an affine $u(1)_4$ current algebra at level four, which is necessary to enhance the $\mathcal N=1$ superconformal algebra to an $\mathcal N=2$ superconformal algebra with central charge 24. The authors of \cite{extN=2} explicitly construct this algebra for the case of the theory with $\Lambda= A_1^{24}$, using the fact that if $j(z)$ is a dimension 1 operator in $\mathcal H_+$, its OPE with any of the twist fields obeys as
\be
j(z) \sigma^a (w) \sim q^{ab} {\sigma^b(w)\over (z-w)}+ \ldots;
\ee
i.e., the twist fields have charge with respect to the dimension one currents in $\mathcal H_+$. This allows one to build an $\mathcal N=2$ superconformal algebra with the correct OPEs.

For the the theory associated with $\Lambda = A_1^{24}$, this is as far as one can go in terms of enhanced supersymmetry, since the orbifold leaves invariant 24 non-interacting dimension one primaries. This is equivalent to the fact that the $\mathcal N=2$ theory for $\Lambda = A_1^{24}$ is an extremal $\mathcal N=2$ theory with $m=4$, whereas the $\mathcal N=2$ theories for all of the other (orbifolded) Niemeier lattices are non-extremal.  However, in this work we will consider the theory with $\Lambda = A_2^{12}$, in which case one can use the invariant primaries to construct an affine $su(2)_4$ current algebra at level 4, which enhances the chiral algebra to an $\mathcal N=4$ superconformal algebra, and that, in fact, this theory satisfies the constraints to be an extremal (chiral) $\mathcal N=4$ superconformal theory with $m=4.$

\section{An extremal chiral $\mathcal N=4$ theory} \label{sec:theory}
First we provide some details about the Niemeier lattice with root system $A_2^{12}$. The lattice is generated by a union of 12 copies of the $A_2$ root system together with additional points which can be written in terms of ``glue vectors". We will call $(x_i,y_i)$ the coordinates in the plane of the $i^{th}$ $A_2$ root system.  A single root system has a basis of simple roots given by
\be
a_1= (\sqrt 2,0)
\ee
and
\be
a_2={1\over \sqrt 2}  (-1,\sqrt 3).
\ee
Then we have  a 24-dimensional basis of simple roots for the full lattice given by
\bea\nonumber
f_1&=&(\sqrt 2, 0,0,0\ldots, 0,0),~h_1={1\over \sqrt 2}(-1,\sqrt 3,0,0\ldots,0,0)\\\nonumber
f_2&=&(0,0,\sqrt 2, 0,\ldots, 0,0),~h_2={1\over \sqrt 2} (0,0,-1,\sqrt 3,\ldots,0,0)\\\nonumber
&\ldots&\\\nonumber
f_{12}&=&(0,0,0,0\ldots,\sqrt 2, 0),~h_{12}={1\over \sqrt 2}(0,0,0,0\ldots,-1,\sqrt 3).
\eea

It is necessary to add glue vectors to the root lattice in order to make the full lattice self-dual. Thus, the glue vectors can naturally be specified in terms of twelve components, each of which lives in the dual lattice of the $i$th $A_2$ root system. There is a specific set of 729 such glue vectors $g_w$ which make up a ``glue code" such that the full lattice is defined as
\be\label{eq:lat1}
\Lambda_{A_2^{12}}=\left\{\sum_{i=1}^{12} (m_i f_i + n_i h_i) + \sum_{w} n_w g_w\Big| m_i, n_i, n_w \in \mathbb Z;~  \sum_w n_w= 0\right \}.
\ee
 More details on the specification of the glue vectors and, in particular, their relation to the ternary Golay code and the Mathieu group $M_{12}$ can be found in appendix \ref{app:glue}.

One can also represent the lattice as an unrestricted sum over 24 basis vectors; it turns out this will be more useful for computation. A full (minimal) basis for the lattice which will be convenient for our purposes can be given in terms of 24 vectors, 18 of which are root vectors, and 6 of which arise from ``glue":
\be\label{eq:lat2}
\Lambda_{A_2^{12}}=\left\{\sum_{i=1}^{12} m_i f_i + \sum_{i=1}^{6}n_i h_i + \sum_{i=1}^6 \ell_i v_i \Big| m_i, n_i, \ell_i \in \mathbb Z\right \}.
\ee
where, defining
\be
\alpha=\left ({1\over \sqrt 2},{1\over \sqrt 6}\right ),
\ee
the additional vectors can be written as
\bea\nonumber
v_1&=&(0, \a,\a,\a,\a,\a,\alpha,0,0,0,0,0)\\\nonumber
v_2&=&(-\a,0, \a,-\a,-\a,\a,0,\a,0,0,0,0)\\\nonumber
v_3&=&(-\a,\a,0,\a,-\a,-\a,0,0,\a,0,0,0)\\\nonumber
v_4&=&(-\a,-\a,\a,0,\a,-\a,0,0,0,\a,0,0)\\\nonumber
v_5&=&(-\a,-\a,-\a,\a,0,\a,0,0,0,0,\a,0)\\
v_6&=&(-\a,\a,-\a,-\a,\a,0,0,0,0,0,0,\a).
\eea
Note that each component of $v_i$ is a vector in the plane of the $i$th root system; i.e., ``0" means the two-dimensional zero vector. One can check that the other six root vectors can be written in terms of linear combinations of these basis vectors with coefficients in $\mathbb Z$. It is possible to check that the representations of the lattice in equation (\ref{eq:lat1}) and (\ref{eq:lat2}) are equivalent using, e.g., properties of the ternary Golay code discussed in appendix \ref{app:glue}.

The full theta function of the lattice, which counts the number of lattice vectors of each length, is given by
\be
\Theta_{A_2^{12}}(\tau)=E_4(\tau)^3 -{81\over 32}\theta_2(\tau)^8\theta_3(\tau)^8\theta_4(\tau)^8.
\ee
The partition function of chiral bosons on this lattice is then 
\be
{\Theta_{A_2^{12}}(\tau)\over \eta(\tau)^{24}}= J(\tau) + 96,
\ee
where the entrance of $1/\eta(\tau)^{24}$ accounts for the 24 dimension one currents $\partial x_i$, and the Virasoro descendents of all of the primaries.

From the representation of the lattice given in (\ref{eq:lat1}), it is easy to understand its symmetry group. The automorphism group of the glue code is the sporadic group $M_{12}$. The automorphism group of the $A_2$ root system consists in the Weyl group, which acts as reflections of the roots, and a $\mathbb Z_2$ action which exchanges the simple roots $a_1$ and $a_2$. Thus, the full automorphism group of the lattice modded out by the Weyl group is the sporadic group $2.M_{12}$, where the 2 indicates a nontrivial double cover of $M_{12}$ \cite{ConwaySloane}. This will be the symmetry group of the (untwisted) bosonic conformal field theory associated to this lattice.

\subsection{Current algebra}
We would like to understand the structure of the lattice CFT under the $\mathbb Z_2$ orbifold which acts as $x_i \to -x_i$, and in particular, to compute the partition function in the Ramond sector to verify that it is equal to the extremal elliptic genus with $m=4$.
First we will analyze the invariant and anti-invariant states under the $\mathbb Z_2$ action. It turns out we will only need to understand the behavior of the dimension one currents in order to fix the partition function of this theory in the Ramond sector. There are eight dimension one currents associated with each  copy of the $A_2$ root system. We write them as
\bea\nonumber
J^{i,-}_x &=& i:\partial x^i:,~ J^{i,-}_y\simeq i:\partial y^i:\\\nonumber
J^{+}_{\lambda_i} &=&{1\over \sqrt 2} \left ( :e^{i \lambda_i\cdot \vec x}:\sigma_{\lambda_i} + :e^{-i \lambda_i\cdot \vec x}:\sigma_{-\lambda_i}\right )\\\nonumber
J^{-}_{\lambda_i} &=&{1\over \sqrt 2i} \left ( :e^{i \lambda_i\cdot \vec x}:\sigma_{\lambda_i} - :e^{-i \lambda_i\cdot \vec x}:\sigma_{-\lambda_i}\right )
\eea
where $\lambda_i= f_i, h_i, f_i+h_i$ denotes one of the three positive roots of the $i$th $A_2$ root system, $\vec x= (x_1,y_1,\ldots,x_{12},y_{12})$, and $::$ denotes the usual normal ordering.
The $\sigma_{\lambda}$s are cocycle factors needed to ensure mutually local  OPEs of these vertex operators. With these factors included, together these currents generate an affine $\widehat{su(3)}_1$ current algebra at level one.

Here we have labeled the currents with a plus or minus sign based on their eigenvalue under the $\mathbb Z_2$ orbifold action. Therefore, we see that there are three invariant currents in the NS sector: $J^{+}_{f_i}, J^{+}_{h_i}$, and $J^{+}_{f_i+h_i}$. Choosing a particular copy of $A_2$ root system, say $i=1$, and redefining the three $\mathbb Z_2$-invariant currents as 
\be
J_1= 2 J^{+}_{f_1+ h_1},~J_2= 2J^{+}_{h_1},~J_3= 2J^{+}_{f_1},
\ee
we see that their OPE satisfies
\be
J_i(z) J_j(w)\sim {4 \delta^{ij}\over (z-w)^2} +i \sqrt 2 \epsilon^{ijk} {J_k(w)\over (z-w)} + reg.
\ee
These are precisely the OPEs of an $\widehat{su(2)}_4$ current algebra. In order to verify this, one needs the following properties of the cocycle factors
\be
\sigma_x\sigma_y = (-1)^{x\cdot y}\sigma_y \sigma_x = \epsilon(x,y)\sigma_{x+y}
\ee
and 
\be
\epsilon(x,y)\epsilon(x+y,z)=\epsilon(x,y+z)\epsilon(y,z).
\ee
We choose a gauge where $\sigma_0=1, \epsilon(x,0)=\epsilon(0,x)=1, \epsilon(x,-x)=-1,$ and $\epsilon(f_1,h_1)=i$. For more discussion of these cocycle operators, see, e.g., \cite{DGM}.

Now we would like to compute the charges of the Ramond sector currents which are eigenstates of a suitably chosen Cartan element of this $su(2)$. Let's select $J_3$ and compute its OPE with the 60 Ramond sector currents. First, it's obvious that for $i\neq 1$, the OPE will be completely regular since the currents are composed of fields which commute with those in $J_3$. Second, it is clear for the same reason that the OPE of $J_3$ with $J^{1,-}_{y}$ will be regular for the same reason. So we need to compute the OPE of $J_3$ with the four Ramond sector currents $J^{1,-}_x, J^{-}_{f_1}, J^{-}_{h_1}, J^{-}_{f_1+h_1}$.

After some work, we find
\be
J_3(z) J^{1,-}_x(w)\sim -2\sqrt 2i { J^{-}_{f_1}(w)\over (z-w)},~~ J_3(z) J^{-}_{f_1}(w) \sim -2\sqrt 2i {J^{1,-}_x(w)\over (z-w)}
\ee
and
\be
J_3(z)J^{-}_{h_1}(w)\sim i \sqrt 2 {J^{-}_{f_1 + h_1}(w)\over (z-w)},~~J_3(z)J^{-}_{f_1 + h_1}(w)\sim i \sqrt 2 {J^{-}_{h_1}(w)\over (z-w)}.
\ee

We see that by taking suitable linear combinations of $J^{1,-}_x$ and $J^{-}_{f_1}$ we get one charge 2 state and one charge $-2$ state, and similarly, linear combinations of $J^{-}_{h_1}$ and $J^{-}_{f_1 + h_1}$ give states of charge $\pm 1$. This fixes the form of the elliptic genus in the Ramond sector to be
\be
{1\over y^4} + {1\over y^2} + 56 + y^2 + y^4 + O(q),
\ee
where the charge grading is given by $2J_3$ as is usual for an $\mathcal N=4$ elliptic genus. With these leading coefficients, the entire Ramond sector partition function is fixed given that it is a weight 0 index 4 weak Jacobi form.\footnote{Strictly speaking, one must also prove the existence of an $\mathcal N=2$ spectral flow symmetry. This amounts to showing that there are two chiral generators $\varepsilon_\pm$ of dimension $\Delta=4$ and $ U(1)$ charges $J_0=\pm 8$ (of an $\mathcal N=2$)\cite{Distler}. We can show this using the same argument in \cite{extN=2}. The $\varepsilon_\pm$ generators they construct are also present in our lattice theory and thus there also exists an $\mathcal N=2$ spectral flow, which, with the relation   $J_0= 2J_3$, is all that is needed to prove the $\mathcal N=4$ partition function is a weak Jacobi form of the correct weight and index.} Therefore we see this reproduces the extremal $\mathcal N=4$ elliptic genus defined in \cite{GGKMO} as the partition function of the Ramond sector in this chiral theory.

\subsection{Superconformal algebra}
In addition to the stress tensor and $su(2)$ current algebra, the $\mathcal N=4$ SCA has four supercurrents $G^\pm_{1,2}$. These transform as doublets under the $su(2)$ with OPEs
\bea\nonumber
J_i(z) G^+_{a}(w) & \sim & -{1\over \sqrt 2 (z-w)} \sigma^i_{ab}G^+_b(w)\\
J_i(z) G^-_{a}(w) & \sim &  {1\over \sqrt 2 (z-w)} (\sigma^i_{ab})^*G^-_b(w),\label{eq:JOPE}
\eea
and their OPEs with each other are
\bea\nonumber
G^+_a(z) G^-_b(w)&\sim& \delta_{ab} \left ({16\over (z-w)^3} + {2T(w)\over z-w}\right ) - \sqrt 2 \sigma_{ab}^i \left ({2 J_i(w)\over (z-w)^2} + {\partial J_i(w) \over z-w}\right )\\
G^+_a(z) G^+_b(w)&\sim&G^-_a(z) G^-_b(w)\sim 0.\label{eq:GOPE}
\eea

We want to construct four dimension-3/2 supercurrents, $G^\pm_{1,2}$ that precisely satisfy the OPEs of the $\mathcal N=4$ superconformal algebra. We need to check the OPEs of the currents with each other and with the $su(2)$ currents such that equations (\ref{eq:JOPE}) and (\ref{eq:GOPE}) hold, and the OPEs with all other dimension one currents are regular. As we have discussed, there are 4096 dimension 3/2 twist fields which form a basis for the dimension 3/2 operators. We will construct $G^\pm_{1,2}$ by taking linear combinations of these twist fields. 

First, note that equations (\ref{eq:JOPE}) show that the four supercurrents are all eigenvectors with respect to $J_3(z)$.  We can take a basis of the 4096 $\sigma^a(z)$ twist fields that is diagonal with respect to $J_3(z)$. There will be 2048 states with positive and negative eigenvalues, which we will write as $\sigma^{+,i}(z)$ and $\sigma^{-,i}(z)$, respectively. We can construct four supercurrents which are eigenstates with respect to $J_3(z)$ and manifestly satisfy equations (\ref{eq:JOPE}) as,
\bea\nonumber
G^\pm_1(z) =  \sum_{i=1}^{2048} c_{\mp, i}^1\sigma^{\mp,i}(z)\\
G^\pm_2(z) =  \sum_{i=1}^{2048} c_{\pm, i}^2 \sigma^{\pm,i}(z),
\eea
where $c_{\pm,i}^{1,2}$ are complex coefficients. The OPEs of equation \ref{eq:JOPE} imply they  satisfy 
\be
c^2_{\pm,i}=\mp c^1_{\mp,i},
\ee
where we have defined the $\sigma^{\pm,i}$ such that $J_1\sigma^{\pm,i} \sim \sigma^{\mp,i}$.
 Here we see that $G^-_1(z)$ and $G^+_2(z)$ have $J_3$ eigenvalue $+1$ and $G^+_1(z)$ and $G^-_2(z)$ have $J_3$ eigenvalue $-1$.

An easier way to verify the OPEs will be to work with expectation values. Define the following two twisted sector ground states to be the ones with either all positive or all negative eigenvalues with respect to the 12 $U(1)$ currents $J_{f_i}$:
\be
||\pm \rangle \equiv |\pm \pm\pm\pm\pm\pm\pm\pm\pm\pm\pm\pm\rangle.
\ee
We claim the following combinations of twisted sector ground states form supercharges which satisfy the OPEs of equation (\ref{eq:GOPE})
\be
G_2^\pm = \prod_{n=2}^{12} (1\pm i J_1^n)||\pm \rangle
\ee
for $a=b=2$. Acting with the $su(2)$ generators and using (\ref{eq:JOPE}) can then generate the other OPEs including $G_1^\pm$.
Here it is convenient to define $J_1^n=J_{f_n+h_n}, J_2^n= J_{h_n}, J_3^n= J_{f_n}$ as $su(2)$ generators associated to the $n$ other $A_2$ root systems.
 Note that $J_1^n ||\pm\rangle$ yields the state $|\pm \pm \ldots \mp \ldots \pm\rangle$ where the $n$th eigenvalue flips sign. With these definitions, it is not difficult to verify that the following expectation values hold,
\bea\nonumber
\langle G_2^-|J_3^i| G_2^+\rangle &\sim& \delta^{i1}\\\nonumber
\langle G_2^-|J_{1,2}^i| G_2^+\rangle &\sim& 0\\\nonumber
\langle G_1^-|J_3^i| G_2^+\rangle &\sim& 0\\\nonumber
\langle G_1^-|J_{1,2}^i| G_2^+\rangle &\sim& \delta^{i1},
\eea
which in turn imply the OPEs of equation (\ref{eq:GOPE}) after proper normalization. In addition, one has to check for decoupling of dimension two operators besides the stress tensor. This is discussed in appendix \ref{app:OPE}.

Finally, we point out that this choice of $su(2)$ current algebra singles out one of the $A_2$ root systems, breaking the automorphism group of the theory from $2.M_{12}$ to $2\times M_{11}$, where $M_{11}$ is the subgroup of $M_{12}$ which stabilizes a point in the 12-dimensional permutation represenation.
\section{Discussion}\label{conclusion}
In this section, we mention a few natural questions given the growing number of explicit extremal (chiral) CFTs which have been constructed.
\begin{itemize}
\item The first and perhaps most obvious question is: are there any other examples of $\mathcal N=2$ or $\mathcal N=4$ theories with an extremal elliptic genus according to \cite{GGKMO}? Are there any non-chiral examples (that don't admit holomorphic factorization)?
\item Is there a suitable modification to the definition of \cite{GGKMO} that loosens the restrictions on states with negative polarity and generalizes to arbitrarily high central charge? Are there families of theories we can construct which satisfy this?
\item Some theories with $\mathcal N=4$ superconformal symmetry have vanishing elliptic genus due to right-moving fermion zero modes. An example of this is given by the elliptic genus of $T^4$ and its symmetric products $(T^4)^N/S_N$\cite{MMS}. Another example where an elliptic genus vanishes is theories with large $\mathcal N=4$ superconformal symmetry\cite{LargeN=4}. However, one can define an index for each of these theories that does not vanish\cite{MMS,GMMS}.  One can try to impose extremal constraints on these functions and see if there are theories which satisfy such constraints.\footnote{Or, if one considers chiral theories with large $\mathcal N=4$ superconformal symmetry, one could impose constraints on its holomorphic partition function. It does not seem like any of the $\mathbb Z_2$ orbifold theories of the Niemeier lattices gives rise to a theory with large $\mathcal N=4$ symmetry since this algebra necessarily includes dimension 1/2 currents. This does not preclude the possibility of a more complicated orbifold yielding such a theory.}
\item  A natural thing to consider in 2d chiral conformal field theories with a large symmetry group $G$ is a ``twined" version of the partition function, with the insertion of some element $g \in G$ in the trace:
\be
Z_g(\t,z) = \tr(g y^{2J_3} q^{L_0-c/24}).
\ee
In the monster theory, these functions are called McKay-Thompson series, and they have interesting modular properties under congruence subgroups of $SL(2,\mathbb Z)$. It may be interesting to study these functions for the theories with $\mathcal N=2,4$ superconformal symmetry and central charge 24, similar to what was done in \cite{M5} for the $c=12$ extremal theories, where the functions were shown to have some interesting mathematical properties.
\item It is natural to organize the massive states in this theory and that of \cite{extN=2} into  vector-valued mock modular forms\cite{Zwegers2008,DMZ}, similar to \cite{M5}. Recently there have been a number of connections uncovered between mock modular forms and finite groups\cite{UMNL}. It may be interesting to investigate these extremal SCFTs and their ``twining" function further from this point of view, or that of Rademacher sums \cite{Farey,ChengDunc}.

\item All extremal CFTs found thus far have states which form representations of large discrete sporadic symmetry groups. These sporadic groups are mathematically interesting because of their connections with unimodular lattices and error-correcting codes\cite{ConwaySloane}. It would be interesting to find a CFT interpretation for these error-correcting codes, and see if these connections manifest in theories with large central charge and semi-classical $AdS_3$ gravity duals.

\item It could be interesting to study the superconformal theories of the $\mathbb Z_2$ orbifolds of the 21 other Niemeier lattices. They will not be extremal with respect to $\mathcal N=2,4$ superconformal algebras, but they may be extremal with respect to some extended chiral algebra, or have other interesting mathematical properties. It would be satisfying if they were somehow related to the moonshine of \cite{UMNL}.

\item  The authors of \cite{PPV} studied an eight-dimensional compactification of the heterotic string where the left-movers were compactified on a $\mathbb Z_2$ orbifold of $\Lambda_{Leech}$. They were able to interpret some interesting properties of the McKay-Thompson series of the monster theory as symmetries acting on two-dimensional spacetime BPS states. It may be interesting to study this for the supersymmetric theories of \cite{DGH}, \cite{extN=2}, or the theory constructed in this paper. In do this, one will likely need to overcome some difficulties in defining a heterotic compactification including this nonlocal orbifold of the left-movers. 

\end{itemize}

\centerline{\bf{Acknowledgements}}
I would like to thank Nathan Benjamin, Miranda Cheng, Ethan Dyer, Shamit Kachru, Natalie Paquette for discussions. I am supported by a Harvard University Golub fellowship in the physical sciences.

\appendix

\section{Useful formulas}\label{app:formulas}

\subsection{Modular and Jacobi forms}
We start by defining the Dedekind eta function,
\be
\eta(\tau) = q^{1/24} \prod_{n=1}^\infty (1-q^n).
\ee
We define the {\em Jacobi theta functions} $\th_i(\t,z)$ as follows for $q=e(\t)$ and $y=e(z)$:
\begin{align}	\th_1(\t,z)
	&= -i q^{1/8} y^{1/2} \prod_{n=1}^\infty (1-q^n) (1-y q^n) (1-y^{-1} q^{n-1})\,,\\
	\th_2(\t,z)
	&=  q^{1/8} y^{1/2} \prod_{n=1}^\infty (1-q^n) (1+y q^n) (1+y^{-1} q^{n-1})\,,\\
	\th_3(\t,z)
	&=  \prod_{n=1}^\infty (1-q^n) (1+y \,q^{n-1/2}) (1+y^{-1} q^{n-1/2})\,,\\
	\th_4(\t,z)
	&=  \prod_{n=1}^\infty (1-q^n) (1-y \,q^{n-1/2}) (1-y^{-1} q^{n-1/2})\,.
\end{align}
Using these functions, we can defined the generators of the ring of weak Jacobi forms which are used in the text. The weight four Eisenstein series $E_4$ can be written as
\be
E_4(\tau) = 1 + 240 \sum_{n=1}^\infty {n^3 q^n\over 1-q^n},
\ee
and the weight six Eisenstein series $E_6$ is
\be
E_6(\tau)= 1 -504 \sum_{n=1}^\infty {n^5 q^n\over 1-q^n}.
\ee
The remaining generators can be written as
\be
\varphi_{0,1}(\t,z) = 4\left ( {\th_2(\t,z)^2\over \th_2(\t,0)^2}+{\th_3(\t,z)^2\over \th_3(\t,0)^2}+{\th_4(\t,z)^2\over \th_4(\t,0)^2}\right )
\ee
and 
\be
\varphi_{-2,1}(\t,z)= {\th_1(\t,z)^2\over \eta(\t)^6}.
\ee

\subsection{$\mathcal N=4$ superconformal characters}
Recall (cf. \cite{Eguchi1987}) that the  ${\cal N}=4$ superconformal algebra contains subalgebras isomorphic to the affine $\SU(2)$ and Virasoro Lie algebras. In a unitary representation the former of these acts with level $m$, for some integer $m\geq1$, and the latter with central charge $c=6m$.  

The unitary irreducible highest weight representations $v^{\cN=4}_{m;h,j}$ are labeled by the eigenvalues of $L_0$ and $\frac{1}{2}J_0^3$ acting on the highest weight state, which we denote by $h$ and $j$, respectively. Cf. \cite{Eguchi1988,Eguchi1988a}.
The superconformal algebra has two types of highest weight Ramond sector representations: the {\em massless} (or {\em BPS}) representations with $h=\frac{c}{24}=\frac{m}{4}$ and $j\in\{ 0,\frac{1}{2},\cdots,\frac{m}{2}\}$, and the {\em massive} (or {\em non-BPS}) representations with $h > \frac{m}{4}$ and $j\in\{ \frac{1}{2},1,\cdots, \frac{m}{2}\}$. We will define their graded characters as
\be
{\rm ch}^{\cN=4}_{m;h,j}(\t,z) = \tr_{v^{\cN=4}_{m;h,j}} \left( (-1)^{J_0^3}y^{J_0^3} q^{L_0-c/24}\right).
\ee

\subsection{$\mathcal N=2$ superconformal characters}

For the SCA with central charge $c= 3(2\ll+1) = 3\hat c$, the unitary irreducible highest weight representations $ v^{\cN=2}_{\ll;h,Q}$ are labeled by the two quantum numbers $h$ and $Q$ which are the eigenvalues of $L_0$ and $J_0$, respectively, when acting on the highest weight state \cite{Dobrev:1986hq,Kiritsis:1986rv}.
Just as in the $\cN=4$ case, 
there are two types of Ramond sector highest weight representations: the {\em massless} (or {\em BPS}) representations with $h=\frac{c}{24} = \frac{\hat c}{8}$ and $Q\in\{ -\frac{\hat c}{2}+1,-\frac{\hat c}{2}+2,\dots,\frac{\hat c}{2}-1,\frac{\hat c}{2}\}$, and the {\em massive} (or {\em non-BPS}) representations with $h > \frac{\hat c}{8}$ and $Q\in\{ -\frac{\hat c}{2}+1,-\frac{\hat c}{2}+2,\dots,\frac{\hat c}{2}-2,\frac{\hat c}{2}-1,\frac{\hat c}{2}\}$, $Q\neq 0$. From now on we will concentrate on the case when $\ll$ is half-integral, and hence $\hat c$ and $c$ are even. 
We write the graded characters as
\be
{\rm ch}^{\cN=2}_{\ll;h,Q}(\t,z) = \tr_{v^{\cN=2}_{\ll;h,Q}} \left( (-1)^{J_0^3}y^{\til J_0^3} q^{L_0-c/24}\right).
\ee
See,e.g., \cite{M5} for explicit formulas for the characters discussed in this section and the next one.

\section{Extremal elliptic genera and their character expansions}\label{app:functions}
Using the characters  discussed in the previous section, we decompose the first five examples of extremal $\mathcal N=2$ and $\mathcal N=4$ elliptic genera defined in \cite{GGKMO}. When possible, we point out cases for which a SCFT with such elliptic genus (or chiral partition function) has been constructed, and comment on their symmetry groups. It is the hope that, given the relationship between known extremal CFTs and discrete sporadic groups, including those connected to the monster \cite{FLM} and Conway \cite{Duncan,DM-C} groups, looking at coefficients of these functions may help in discovering more examples of extremal theories and/or in unifying our understanding of extremal theories in general. As mentioned earlier in the text, if there are theories not yet discovered which have extremal elliptic genera for $m=1,3,5$, they will necessarily be non-chiral.
\begin{itemize}
\item $m=1$

In this case the extremal $\mathcal N=2$ and $\mathcal N=4$ elliptic genera are the same and given by
\be
Z^{\mathcal N=2}_{m=1}(\tau,z) =Z^{\mathcal N=4}_{m=1}(\tau,z) =\varphi_{0,1}(\tau,z).
\ee
The decomposition in $\mathcal N=2$ characters is
\be
Z^{\mathcal N=2}_{m=1}(\tau,z)=11{\rm ch}_{{1\over 2};{1\over 4},0}^{\mathcal N=2}(\t,z) -{\rm ch}_{{1\over 2};{1\over 4},1}^{\mathcal N=2}(\t,z) + \sum_{n=1}^\infty A_n {\rm ch}_{{1\over 2};{1\over 4}+n,1}^{\mathcal N=2}(\t,z)
\ee
and the decomposition into $\mathcal N=4$ characters is
\be
Z^{\mathcal N=4}_{m=1}(\tau,z)=10{\rm ch}_{1;{1\over 4},0}^{\mathcal N=4}(\t,z) -{\rm ch}_{1;{1\over 4},{1\over 2}}^{\mathcal N=4}(\t,z) + \sum_{n=1}^\infty \tilde A_n{\rm ch}_{1;{1\over 4}+n,{1\over 2}}^{\mathcal N=4}(\t,z),
\ee
where $A_n= \tilde A_n = \{45, 231, 770, 2277, \ldots\}$ for $n=\{1,2,3, 4, \ldots\}$.

It is interesting to note that the elliptic genus of a K3 surface is $2\varphi_{0,1}(\t,z)$ and could be considered extremal under some definition. The coefficients $A_n$ are the same numbers which were seen to be related to dimensions of irreducible representations of the Mathieu group $M_{24}$ as first noticed by \cite{EOT} when studying the expansion of the K3 elliptic genus in $\mathcal N=4$ superconformal characters.
\item $m=2$

In this case the central charge is 12 and we notice that the extremal $\mathcal N=2$ and $\mathcal N=4$ elliptic genera are again the same function:
\be
Z^{\mathcal N=2}_{m=2}(\tau,z) =Z^{\mathcal N=4}_{m=2}(\tau,z) ={5\over 6} E_4\varphi_{-2,1}^2 + {1\over 6}\varphi_{0,1}^2.
\ee

The character decompositions are 
\be\nonumber
Z^{\mathcal N=2}_{m=2}(\tau,z)=23{\rm ch}^{\mathcal N=2}_{{3\over 2};{1\over 2},0}(\t,z) +{\rm ch}^{\mathcal N=2}_{{3\over 2};{1\over 2},{2}}(\t,z)+ \sum_{n=1}^\infty  \sum_{k=\pm 1,2}A_{n,k} {\rm ch}^{\mathcal N=2}_{{3\over 2};{1\over 2}+n,k}(\t,z)
\ee
where $A_{n,1}=A_{n,-1} = \{770, 13915, 132825+\ldots\}$ and $A_{n,2}=\{231, 5796, 65505+\ldots\}$. Similarly, the $\mathcal N=4$ decompositions are
\be
Z^{\mathcal N=4}_{m=2}(\tau,z) =21{\rm ch}^{\mathcal N=4}_{2;{1\over 2},0}(\t,z) +{\rm ch}^{\mathcal N=4}_{2;{1\over 2},{1}}(\t,z) + \sum_{n=1}^\infty \sum_{k=1}^2 \tilde A_{n,{k\over 2}} {\rm ch}^{\mathcal N=4}_{2;{1\over 2}+n,{k\over 2}}(\t,z)
\ee
where $\tilde A_{n,{1\over 2}} = \{560, 8470, 70576, \ldots\}$ and $\tilde A_{n,1} =\{210, 4444,  42560,\ldots\}$. A chiral theory with this graded partition function and $\mathcal N=2,4$ supersymmetry was constructed in \cite{M5}. The coefficients $A_{n,k}$ and $\tilde A_{n,k}$ are related to dimensions of irreducible representations of the Mathieu groups $M_{23}$ and $M_{22}$, respectively.
\item $m=3$

This case corresponds to central charge 18. At this point, and for all higher values of $c$ where they exist, the extremal $\mathcal N=2$ and $\mathcal N=4$ elliptic genera are distinct functions. For this case we have
\bea\nonumber
Z^{\mathcal N=2}_{m=3}(\tau,z) &=&{13\over 24}\varphi_{-2,1}E_6 + {7\over 16}\varphi_{-2,1}^2\varphi_{0,1}E_4 + {1\over 48}\varphi_{0,1}^3\\\nonumber
&=&35{\rm ch}^{\mathcal N=2}_{{5\over 2}; {3\over 4},0}(\t,z) - {\rm ch}^{\mathcal N=2}_{{5\over 2}; {3\over 4},3}(\t,z)+\sum_{n=1}^\infty  \sum_{k=\pm 1,\pm 2,3}A_{n,k} {\rm ch}^{\mathcal N=2}_{{5\over 2};{3\over 4}+n,k}(\t,z),
\eea
where the first few coefficients are $A_{n, 1} = A_{n,-1} = \{5984, 262140, 5078546,\ldots\}$, $A_{n, 2} = A_{n,-2} = \{2244, 132396, 2920005,\ldots\}$, and $A_{n,3} = \{187, 30261, 911098,\ldots\}$.

The $\mathcal N=4$ form is given by
\bea\nonumber
Z^{\mathcal N=4}_{m=3}(\tau,z) &=&{59\over 108}\varphi_{-2,1}E_6 + {31\over 72}\varphi_{-2,1}^2\varphi_{0,1}E_4 + {5\over 216}\varphi_{0,1}^3\\\nonumber
&=&36{\rm ch}^{\mathcal N=4}_{{3}; {3\over 4},0}(\t,z) - {\rm ch}^{\mathcal N=4}_{{3}; {3\over 4},{3\over 2}}(\t,z)+ \sum_{n=1}^\infty \sum_{k=1}^3 \tilde A_{n,{k\over 2}} {\rm ch}^{\mathcal N=4}_{3;{3\over 4}+n,{k\over 2}}(\t,z)
\eea
where the first few coefficients are $\tilde A_{n, {1\over 2}} = \{3780,131328, \ldots\}$, $\tilde A_{n, {1}} = \{2016, 98118,\ldots\}$, and $\tilde A_{n, {3\over 2}} = \{189, 22267, \ldots\}$. Neither an extremal $\mathcal N=2$ nor $\mathcal N=4$ SCFT has been discovered with this central charge.

\item $m=4$

The $\mathcal N=2$ theory is the case discussed in \cite{extN=2}, and was shown to have $M_{23}$ symmetry. The Jacobi form is
\bea\nonumber
Z^{\mathcal N=2}_{m=4}(\tau,z) &=&{67 \over 144} \varphi_{-2,1}^4E_4^2 + {11\over 27}\varphi_{-2,1}^3\varphi_{0,1}E_6 + {1\over 8}\varphi_{-2,1}^2\varphi_{0,1}^2E_4 + {1\over 432}\varphi_{0,1}^4\\\nonumber
&=&47 {\rm ch}^{\mathcal N=2}_{{7\over 2}; {1},0}(\t,z) + {\rm ch}^{\mathcal N=2}_{{7\over 2}; {1},4}(\t,z)+\sum_{n=1}^\infty  \sum_{k=\pm 1,\pm 2,\pm 3,4}A_{n,k} {\rm ch}^{\mathcal N=2}_{{7\over 2};{1}+n,k}(\t,z)
\eea
where the coefficients are: $A_{n, 1} = A_{n,-1} = \{32890, 2969208 ,\ldots\}$, $A_{n, 2} = A_{n,-2} = \{14168, 1659174,\ldots\}$, $A_{n, 3} = A_{n,-3} = \{2024, 485001,\ldots\}$, and $A_{n,4} = \{23, 61894,\ldots\}$.

The extremal $\mathcal N=4$ Jacobi form is the one discussed in this paper and is given by
\bea\nonumber
Z^{\mathcal N=4}_{m=4}(\tau,z) &=&{271 \over 576} \varphi_{-2,1}^4E_4^2 + {43\over 108}\varphi_{-2,1}^3\varphi_{0,1}E_6 + {37\over 288}\varphi_{-2,1}^2\varphi_{0,1}^2E_4 + {5\over 1728}\varphi_{0,1}^4\\\nonumber
&=&55 {\rm ch}^{\mathcal N=4}_{{4}; {1},0}(\t,z) + {\rm ch}^{\mathcal N=4}_{{4}; {1},2}(\t,z)+ \sum_{n=1}^\infty \sum_{k=1}^4 \tilde A_{n,{k\over 2}} {\rm ch}^{\mathcal N=4}_{4;{1}+n,{k\over 2}}(\t,z),
\eea
where the coefficients are: $\tilde A_{n, {1\over 2}} = \{18876, 1315512, \ldots\}$, $\tilde A_{n, {1}} = \{12045, 1152943,\ldots\}$, $\tilde A_{n, {3\over 2}} = \{1980, 391974, \ldots\}$, and $\tilde A_{n, { 2}} = \{33, 45990, \ldots\}$. These coefficients are related to dimensions of representations of the sporadic group $M_{11}$.

\item $m=5$

The central charge in this case is $c=30$. No such extremal $\mathcal N=2$ or $\mathcal N=4$ theory has been found. In the $\mathcal N=2$ cases, the extremal elliptic genus is,
\bea\nonumber
Z^{\mathcal N=2}_{m=5}(\tau,z) &=&{2975 \over 6912} \varphi_{-2,1}^4\varphi_{0,1}E_4^2 + {1979\over 5184}\varphi_{-2,1}^5E_4E_6 + {835\over 5184}\varphi_{-2,1}^3\varphi_{0,1}^2E_6 + {275\over 10368}\varphi_{-2,1}^2\varphi_{0,1}^3E_4 + {5\over 20376}\varphi_{0,1}^5\\\nonumber
&=&59 {\rm ch}^{\mathcal N=2}_{{9\over 2}; {5\over 4},0}(\t,z) - {\rm ch}^{\mathcal N=2}_{{9\over 2}; {5\over 4},5}(\t,z)+\sum_{n=1}^\infty  \sum_{k=\pm 1,\pm 2,\pm 3,\pm 4,5}A_{n,k} {\rm ch}^{\mathcal N=2}_{{9\over 2};{5\over 4}+n,k}(\t,z),
\eea
where the coefficients are: $A_{n, 1} = A_{n,-1} = \{146566, 24757474 ,\ldots\}$, $A_{n, 2} = A_{n,-2} = \{69426, 14772861,\ldots\}$, $A_{n, 3} = A_{n,-3} = \{13224, 5026222,\ldots\}$, $A_{n, 4} = A_{n,-4} = \{551, 868927,\ldots\}$, and $A_{n,5} = \{0, 57798,\ldots\}$.

Finally, the extremal $\mathcal N=4$ elliptic genus is,
\bea\nonumber
Z^{\mathcal N=4}_{m=5}(\tau,z) &=&{2941 \over 6912} \varphi_{-2,1}^4\varphi_{0,1}E_4^2 + {1999\over 5184}\varphi_{-2,1}^5E_4E_6 + {827\over 5184}\varphi_{-2,1}^3\varphi_{0,1}^2E_6 + {301\over 10368}\varphi_{-2,1}^2\varphi_{0,1}^3E_4 + {7\over 20376}\varphi_{0,1}^5\\\nonumber
&=&78 {\rm ch}^{\mathcal N=4}_{5; {5\over 4},0}(\t,z) - {\rm ch}^{\mathcal N=4}_{{5}; {5\over 4},{5\over 2}}(\t,z)+ \sum_{n=1}^\infty \sum_{k=1}^5 \tilde A_{n,{k\over 2}} {\rm ch}^{\mathcal N=4}_{5;{5\over 4}+n,{k\over 2}}(\t,z),
\eea
where the coefficients are: $\tilde A_{n, {1\over 2}} = \{78078, 10007569, \ldots\}$, $\tilde A_{n, {1}} = \{56056, 9655568,\ldots\}$, $\tilde A_{n, {3\over 2}} = \{12441, 4017663, \ldots\}$,  $\tilde A_{n, {2}} = \{572, 742456, \ldots\}$, and $\tilde A_{n, {5\over 2}} = \{0, 44682, \ldots\}$.
\end{itemize}
At this point we have exhausted all possible extremal $\mathcal N=4$ elliptic genera as defined in \cite{GGKMO}. There are functions satisfying the extremal $\mathcal N=2$ condition also for $m=7,8,11$ and 13, but we don't reproduce them here.

\section{The ternary Golay code and the Mathieu group $M_{12}$}\label{app:glue}
The discussion in this section in large part comes from chapter 3 of \cite{ConwaySloane}. A binary code of length $n$ is a set of binary vectors called codewords with $n$ coordinates taking values in the field $\mathbb F_2^n$ where $\mathbb F_2=\{0,1\}$. Similarly, a $q$-ary code of length $n$ is a set of codewords taking values in the field $\mathbb F_q^n$ where $\mathbb F_q$ is the field of integers mod $q$ and $q= p^a$ for some prime $p$.  For the purposes of error-correction, one wants to choose codewords which are in some sense easy to distinguish from each other in case some errors have occurred. A precise measure of this is what is known as the Hamming distance between two codewords
\be\nonumber
u=(u_1,\ldots, u_n),~~v=(v_1,\ldots, v_n),
\ee
defined to be the number of coordinates where they differ,
\be
d(u,v) = |\{i:u_i \neq v_i\}|.
\ee
The Hamming weight of a vector $u$, denoted as ${\rm wt}(u)$, is the number of nonzero coordinates of $u$; therefore
\be
d(u,v) = {\rm wt}(u-v).
\ee
The minimal distance $d$ of a code is the ``closest distance possible" between two codewords, i.e.
\be
d= \min \{d(u,v): u,v \in C, u\neq v\}.
\ee
A code with minimal distance $d$ has ``packing radius" $\rho$ where
\be
\rho= {1\over 2}(d-1)
\ee
denotes the radius of disjoint ``Hamming spheres" around the codewords; such a code can correct $\rho$ errors.

A linear code is a subspace of $\mathbb F_q^n$, codewords are vectors in this subspace, and the code is closed under vector addition and multiplication by elements of $\mathbb F_q$. The dimension $k$ of a code $C$ is the dimension of this subspace; there are in total $M := q^k$ codewords. In a linear code, the minimal distance is just the minimal nonzero weight of any codeword
\be
d :=\min \{{\rm wt}(u): u\in C, u \neq 0\},
\ee
where ${\rm wt}(u)$ denotes the number of nonzero entries in $u$. For purposes of error-correction, it is desirable to have a small $n$ and large $M$ to increase efficiency, and large $d$ to correct a greater number of errors. A linear code of length $n$, dimension $k$, and minimal distance $d$ is often called an $[n,k,d]$ code. An $[n,k,d]$ code $C$ can be specified by a generator matrix, which is a $k\times n$ matrix such that $C$ consists of all linear combinations of the rows of the matrix with coefficients in $\mathbb F_q$.

Given a code $C$, one can define its dual code, $C^*$, as
\be
C^*=\{x\in \mathbb F_q^n: x\cdot\bar u=0~ \forall ~u \in C\}
\ee
where $\bar u$ denotes conjugation in the field $\mathbb F_q$, that is
\be\nonumber
u \mapsto \bar u = u^p
\ee
when $q=p^a$ for prime $p$. A code where $C=C^*$ is said to be self-dual. A self-dual code necessarily has even $n$ and $k=n/2$.

For linear codes, let $A_i$ denote the number of codewords of weight $i$. Then the weight enumerator of a linear code $C$ is defined to be
\be
W_C(x,y) = \sum_{i=0}^n A_i x^{n-i} y^i
\ee
and counts the number of codewords of each weight.

A cyclic code is a code such that if $c_0c_1\ldots c_{n-1}$ is a codeword, then $c_1c_2\ldots c_{n-1}c_0$ is also a codeword. Representing a codeword $c=c_0c_1\ldots c_{n-1}$ by a polynomial $c(x)= c_0 + c_1x+\ldots c_{n-1}x^{n-1}$, a linear cyclic code can be completely specified by a generator polynomial $g(x)$ which divides $x^n-1$ over $\mathbb F_q$. If $p$ and $n$ are primes, a quadratic residue code of length $n$ over $\mathbb F_p$ is a cyclic code whose generator polynomial has roots $\{\alpha_i: i\neq 0, i = x^2~ {\rm mod} ~n\}$. This code has dimension $(n+1)/2$. An extended quadratic residue code is obtained from a quadratic residue by appending a zero-sum check digit. That is, for each codeword $c_0c_1\ldots c_{n-1}$ one adds an extra digit
\be
c_n = -\sum_{i=0}^{n-1} c_i.
\ee
Then if $n=3~{\rm mod}~4$, the extended code is self dual.

At this point, we are  in position to define the ternary Golay codes. The $[11,6,5]$ ternary Golay code $C_{11}$ is defined as the quadratic residue code of length 11 over $\mathbb F_3$. It is a perfect code, which means that it saturates the ``Hamming bound"; i.e. its packing radius is equal to its covering radius. The (extended) ternary Golay code $C_{12}$ is a $[12,6,6]$ code which is self-dual and obtained from $C_{11}$ by appending a zero-sum check digit. A possible generator matrix is
\be \left( \begin{array}{rrrrrrrrrrrr}
0&1&1&1&1&1&1&0&0&0&0&0 \\
-1&0&1&-1&-1&1&0&1&0&0&0&0 \\
-1&1&0&1&-1&-1&0&0&1&0&0&0 \\
-1&-1&1&0&1&-1&0&0&0&1&0&0 \\
-1&-1&-1&1&0&1&0&0&0&0&1&0 \\
-1&1&-1&-1&1&0&0&0&0&0&0&1  \end{array} \right),\ee
and it has weight enumerator
\be\nonumber
x^{12} + 264x^6y^6 + 440 x^3y^9 + 24 y^{12}.
\ee
The Mathieu group $M_{12}$ is the automorphism group of $C_{12}$, and it acts naturally as a subgroup of the permutation group $S_{12}$ on twelve elements.

\section{Gamma matrix algebra, supercurrents, and OPEs}\label{app:OPE}
We want to understand the nature of the twisted sector ground states in order to construct the supercurrents of the $\mathcal N=4$ algebra. We define the twisted sector ground states as
\be
|a\rangle = \lim_{z\to 0} \sigma^a(z) |0\rangle
\ee
where $|0\rangle$ is the NS sector vacuum. These $2^{12}$ twisted sector ground states form an irreducible representation of the untwisted sector operator algebra. In \cite{DGM} it is shown that a vertex operator corresponding to a lattice vector $\lambda \in \Lambda$ acts on twisted sector ground states as
\be
\lim_{z\to 0}(4z)^{\lambda^2/2}V_{\lambda}(z)|a\rangle = \gamma_\lambda |a\rangle,
\ee
where the nontrivial action is encoded in the cocycle factor $\gamma_\lambda$. In order for the vertex operators to be mutually local, these factors must satisfy
\be\label{eq:gamma}
\gamma_\lambda\gamma_\mu = (-1)^{\lambda \cdot \mu}\gamma_\mu\gamma_\lambda=\epsilon(\lambda,\mu)\gamma_{\lambda+\mu}.
\ee
The nontrivial part of the operator algebra is given by $\Lambda/2\Lambda$ since all vectors in $2\Lambda$ yield commuting vertex operators; in the case of a $\mathbb Z_2$ orbifold of a 24-dimensional lattice, this algebra is a 24-dimensional Clifford algebra.

Now we give an explicit representation of this algebra for the case of $\Lambda=\Lambda_{A_2^{12}}$.  A basis for $\Lambda/2\Lambda$ can be given in terms of the basis for lattice vectors in equation (\ref{eq:lat2}), and one can verify that a choice of gamma matrix algebra satisfying (\ref{eq:gamma})  is
\bea\nonumber
\gamma_{f_1} &=& \sigma_3 \otimes {\bf 1} \otimes \ldots \otimes {\bf 1}\\\nonumber
\gamma_{f_2} &=&{\bf 1} \otimes {\sigma_3} \otimes \ldots \otimes {\bf 1}\\
&\ldots&\\\nonumber
\gamma_{f_{12}} &=&{\bf 1} \otimes {\bf 1} \otimes \ldots \otimes {\sigma_3}\\\nonumber
\gamma_{h_1} &=& \sigma_2 \otimes {\bf 1} \otimes \ldots \otimes {\bf 1}\\\nonumber
\gamma_{h_2} &=&{\bf 1} \otimes {\sigma_2} \otimes \ldots \otimes {\bf 1}\\\nonumber
&\ldots&\\\nonumber
\gamma_{h_{6}} &=&{\bf 1} \otimes {\bf 1} \otimes {\bf 1} \otimes {\bf 1} \otimes{\bf 1} \otimes {\sigma_2} \otimes{\bf 1} \otimes {\bf 1} \otimes{\bf 1} \otimes {\bf 1} \otimes{\bf 1} \otimes {\bf 1} 
\eea
for the root vectors,
and
\bea\nonumber
\gamma_{v_1} &=&{\bf 1} \otimes \sigma_2 \otimes\sigma_2 \otimes\sigma_2 \otimes\sigma_2 \otimes {\sigma_2} \otimes\sigma_2 \otimes {\bf 1} \otimes{\bf 1} \otimes {\bf 1} \otimes{\bf 1} \otimes {\bf 1} \\\nonumber
\gamma_{v_2} &=&{\sigma_2} \otimes {\bf 1} \otimes\sigma_2 \otimes\sigma_2 \otimes\sigma_2 \otimes {\sigma_2} \otimes {\bf 1} \otimes \sigma_2 \otimes{\bf 1} \otimes {\bf 1} \otimes{\bf 1} \otimes {\bf 1} \\\nonumber
\gamma_{v_3} &=&\sigma_2 \otimes\sigma_2\otimes{\bf 1} \otimes \sigma_2 \otimes\sigma_2 \otimes\sigma_2  \otimes {\bf 1} \otimes{\bf 1}\otimes\sigma_2 \otimes {\bf 1} \otimes{\bf 1} \otimes {\bf 1} \\\nonumber
\gamma_{v_4} &=& \sigma_2 \otimes\sigma_2 \otimes\sigma_2 \otimes {\bf 1} \otimes \sigma_2 \otimes {\sigma_2}  \otimes {\bf 1} \otimes{\bf 1} \otimes {\bf 1} \otimes\sigma_2\otimes{\bf 1} \otimes {\bf 1} \\\nonumber
\gamma_{v_5} &=&\sigma_2 \otimes\sigma_2 \otimes\sigma_2 \otimes\sigma_2 \otimes {\bf 1}  \otimes {\sigma_2}  \otimes {\bf 1} \otimes{\bf 1} \otimes {\bf 1} \otimes{\bf 1}\otimes\sigma_2 \otimes {\bf 1} \\
\gamma_{v_6} &=&\sigma_2 \otimes\sigma_2 \otimes\sigma_2 \otimes\sigma_2 \otimes {\sigma_2} \otimes{\bf 1} \otimes  {\bf 1} \otimes{\bf 1} \otimes {\bf 1} \otimes{\bf 1} \otimes {\bf 1}\otimes\sigma_2 
\eea
for the rest of the basis.

We can now use this to verify some of the properties of the supercurrents defined in the text. With this choice of gamma matrix algebra, it is easy to verify the expectation values from section 4. For example
\be
\langle G_2^+|J_3^n|G_2^-\rangle \sim \delta^{n1}
\ee
since terms with positive and negative charge with respect to $J_3^n$ cancel in the sum when $n\neq 1$. One also has to check that no dimension two currents couple to the supercurrents besides the stress tensor. As in \cite{extN=2}, there are three types of dimension two currents of the form: a) $\partial x_i\partial x_j$, b) $e^{i\lambda\cdot \vec x}$ for $\lambda$ the some of two (distinct) root vectors, and c) $e^{i\lambda\cdot \vec x}$ for $\lambda = \sum_{i = 1}^6 a_i v_i$ such that $\lambda^2=4$. The argument for the decoupling of operators of type a) is the same as in \cite{extN=2}. It is possible, given the form of the gamma matrix algebra and the supercurrents defined in the text to explicitly check the decoupling of operators of types b) and c). Note that each $v_i$ itself leads to a dimension two operator and these $v_i$ each individually decouple and commute with each other, implying the decoupling of all such operators of type c).


\begin{thebibliography}{9}



\bibitem{Magoo} 
  O.~Aharony, S.~S.~Gubser, J.~M.~Maldacena, H.~Ooguri and Y.~Oz,
  ``Large N field theories, string theory and gravity,''
  Phys.\ Rept.\  {\bf 323}, 183 (2000)
  [hep-th/9905111].
  
\bibitem{Witten} 
  E.~Witten,
  ``Three-Dimensional Gravity Revisited,''
    arXiv:0706.3359 [hep-th].
  
  \bibitem{hoehn} 
    G.~Hoehn
  ``Conformal Designs based on Vertex Operator Algebras,''
  arXiv:0701626 [math.QA].
  
    \bibitem{FLM} 
  I.~Frenkel, J.~Lepowsky and A.~Meurman,
  ``Vertex Operator Algebras And The Monster,''
  BOSTON, USA: ACADEMIC (1988) 508 P. (PURE AND APPLIED MATHEMATICS, 134)
  
  \bibitem{Duncan}
J. Duncan, ``Super-moonshine for Conway's largest sporadic group," arXiv:math/0502267.


\bibitem{DM-C} 
  J.~F.~R.~Duncan and S.~Mack-Crane,
  ``The Moonshine Module for Conway's Group,''
  SIGMA {\bf 3}, e10 (2015)
  [arXiv:1409.3829 [math.RT]].
  
\bibitem{DGH} 
  L.~J.~Dixon, P.~H.~Ginsparg and J.~A.~Harvey,
  ``Beauty and the Beast: Superconformal Symmetry in a Monster Module,''
  Commun.\ Math.\ Phys.\  {\bf 119}, 221 (1988).
  
  \bibitem{MalWit} 
  A.~Maloney and E.~Witten,
  ``Quantum Gravity Partition Functions in Three Dimensions,''
  JHEP {\bf 1002}, 029 (2010)
  [arXiv:0712.0155 [hep-th]].
  
  \bibitem{yin1} 
  X.~Yin,
  ``On Non-handlebody Instantons in 3D Gravity,''
  JHEP {\bf 0809}, 120 (2008)
  [arXiv:0711.2803 [hep-th]].
  
  \bibitem{yin2} 
  X.~Yin,
  ``Partition Functions of Three-Dimensional Pure Gravity,''
  Commun.\ Num.\ Theor.\ Phys.\  {\bf 2}, 285 (2008)
  [arXiv:0710.2129 [hep-th]].
  
  \bibitem{gaiottoyin} 
  D.~Gaiotto and X.~Yin,
  ``Genus two partition functions of extremal conformal field theories,''
  JHEP {\bf 0708}, 029 (2007)
  [arXiv:0707.3437 [hep-th]].
  
  \bibitem{gab} 
  M.~R.~Gaberdiel,
  ``Constraints on extremal self-dual CFTs,''
  JHEP {\bf 0711}, 087 (2007)
  [arXiv:0707.4073 [hep-th]].
  
  \bibitem{gabetal} 
  M.~R.~Gaberdiel, C.~A.~Keller and R.~Volpato,
  ``Genus Two Partition Functions of Chiral Conformal Field Theories,''
  Commun.\ Num.\ Theor.\ Phys.\  {\bf 4}, 295 (2010)
  [arXiv:1002.3371 [hep-th]].
  
  \bibitem{gabkel} 
  M.~R.~Gaberdiel and C.~A.~Keller,
  ``Modular differential equations and null vectors,''
  JHEP {\bf 0809}, 079 (2008)
  [arXiv:0804.0489 [hep-th]].
  
  \bibitem{gaiotto} 
  D.~Gaiotto,
  ``Monster symmetry and Extremal CFTs,''
  arXiv:0801.0988 [hep-th].
  
  \bibitem{LSS} 
  W.~Li, W.~Song and A.~Strominger,
  ``Chiral Gravity in Three Dimensions,''
  JHEP {\bf 0804}, 082 (2008)
  [arXiv:0801.4566 [hep-th]].

  
  \bibitem{GGKMO} 
  M.~R.~Gaberdiel, S.~Gukov, C.~A.~Keller, G.~W.~Moore and H.~Ooguri,
  ``Extremal N=(2,2) 2D Conformal Field Theories and Constraints of Modularity,''
  Commun.\ Num.\ Theor.\ Phys.\  {\bf 2}, 743 (2008)
  [arXiv:0805.4216 [hep-th]].
  
  \bibitem{KYY} 
  T.~Kawai, Y.~Yamada and S.~K.~Yang,
  ``Elliptic genera and N=2 superconformal field theory,''
  Nucl.\ Phys.\ B {\bf 414}, 191 (1994)
  [hep-th/9306096].
  
  \bibitem{Farey} 
  R.~Dijkgraaf, J.~M.~Maldacena, G.~W.~Moore and E.~P.~Verlinde,
  ``A Black hole Farey tail,''
  hep-th/0005003.
    
  \bibitem{M5} 
  M.~C.~N.~Cheng, X.~Dong, J.~F.~R.~Duncan, S.~Harrison, S.~Kachru and T.~Wrase,
  ``Mock Modular Mathieu Moonshine Modules,''
  Research in the Mathematical Sciences (2015) 2:13
  [arXiv:1406.5502 [hep-th]].
  
    \bibitem{extN=2} 
  N.~Benjamin, E.~Dyer, A.~L.~Fitzpatrick and S.~Kachru,
  ``An extremal ${\mathcal{N}}=2$ superconformal field theory,''
  J.\ Phys.\ A {\bf 48}, no. 49, 495401 (2015)
  [arXiv:1507.00004 [hep-th]].
  
  \bibitem{spin7} 
  N.~Benjamin, S.~M.~Harrison, S.~Kachru, N.~M.~Paquette and D.~Whalen,
  ``On the elliptic genera of manifolds of Spin(7) holonomy,''
  arXiv:1412.2804 [hep-th].
  
\bibitem{exceptional} 
  M.~C.~N.~Cheng, S.~M.~Harrison, S.~Kachru and D.~Whalen,
  ``Exceptional Algebra and Sporadic Groups at c=12,''
  arXiv:1503.07219 [hep-th].
  
  \bibitem{EZ}
  M~ Eichler and D.~Zagier. 1985. \emph{The theory of Jacobi forms.} Birkh\"auser.
  

  
  \bibitem{schell} 
  A.~N.~Schellekens,
  ``Meromorphic C = 24 conformal field theories,''
  Commun.\ Math.\ Phys.\  {\bf 153}, 159 (1993)
  doi:10.1007/BF02099044
  [hep-th/9205072].
  
  \bibitem{mont} 
  P.~S.~Montague,
  ``Orbifold constructions and the classification of selfdual c = 24 conformal field theories,''
  Nucl.\ Phys.\ B {\bf 428}, 233 (1994)
  doi:10.1016/0550-3213(94)90201-1
  [hep-th/9403088].
  

  

  
  \bibitem{ConwaySloane}
 J.~H.~Conway and N.~J.~A.~Sloane,
 ``Sphere packings, lattices and groups,''
 Springer Science \& Business Media, {\bf 290} (2013)

  \bibitem{DGM} 
  L.~Dolan, P.~Goddard and P.~Montague,
  ``Conformal Field Theory of Twisted Vertex Operators,''
  Nucl.\ Phys.\ B {\bf 338}, 529 (1990).
  
  \bibitem{Distler} 
  J.~Distler,
  ``Notes on N=2 sigma models,''
  In *Trieste 1992, Proceedings, String theory and quantum gravity '92* 234-256, and Princeton U. - PUPT-1365 (92/12,rec.Feb.93) 26 p.
  [hep-th/9212062].
  
    \bibitem{MMS} 
  J.~M.~Maldacena, G.~W.~Moore and A.~Strominger,
  ``Counting BPS black holes in toroidal Type II string theory,''
  hep-th/9903163.
  
  \bibitem{LargeN=4} 
  A.~Sevrin, W.~Troost and A.~Van Proeyen,
  ``Superconformal Algebras in Two-Dimensions with N=4,''
  Phys.\ Lett.\ B {\bf 208}, 447 (1988).
  
  \bibitem{GMMS} 
  S.~Gukov, E.~Martinec, G.~W.~Moore and A.~Strominger,
  ``An Index for 2-D field theories with large N = 4 superconformal symmetry,''
  hep-th/0404023.
  



\bibitem{Zwegers2008}
S.~Zwegers,
  ``Mock Theta Functions,''
  arXiv:0807.4834 [math.NT].
  
       \bibitem{DMZ}
A.~Dabholkar, S.~Murthy, and D.~Zagier, ``{Quantum Black Holes, Wall Crossing,
  and Mock Modular Forms},''
\href{http://arxiv.org/abs/1208.4074}{{\tt arXiv:1208.4074 [hep-th]}}.

  \bibitem{UMNL} 
  M.~C.~N.~Cheng, J.~F.~R.~Duncan and J.~A.~Harvey,
  ``Umbral Moonshine and the Niemeier Lattices,''
  arXiv:1307.5793 [math.RT].
  
  \bibitem{ChengDunc} 
  M.~C.~N.~Cheng and J.~F.~R.~Duncan,
  ``On Rademacher Sums, the Largest Mathieu Group, and the Holographic Modularity of Moonshine,''
  Commun.\ Num.\ Theor.\ Phys.\  {\bf 6}, 697 (2012)
  [arXiv:1110.3859 [math.RT]].
  
  \bibitem{PPV} 
  N.~M.~Paquette, D.~Persson and R.~Volpato,
  ``Monstrous BPS-Algebras and the Superstring Origin of Moonshine,''
  arXiv:1601.05412 [hep-th].
  
    \bibitem{Eguchi1987}
 T.~Eguchi and A.~Taormina,
  ``Unitary representations of the {$N=4$} superconformal algebra,''
  Phys.\ Lett.\ B {\bf 196}, 1 (1987).
  \bibitem{Eguchi1988}
 T.~Eguchi and A.~Taormina,
  ``Character Formulas for the $N=4$ Superconformal Algebra,''
  Phys.\ Lett.\ B {\bf 200}, 315 (1988).

  \bibitem{Eguchi1988a}
T.~Eguchi and A.~Taormina,
  ``On the Unitary Representations of $N=2$ and $N=4$ Superconformal Algebras,''
  Phys.\ Lett.\ B {\bf 210}, 125 (1988).
  
  \bibitem{Dobrev:1986hq}
  V.~K.~Dobrev,
  ``Characters of the Unitarizable Highest Weight Modules Over the $N=2$ Superconformal Algebras,''
  Phys.\ Lett.\ B {\bf 186}, 43 (1987).

\bibitem{Kiritsis:1986rv}
  E.~Kiritsis,
  ``Character Formulae and the Structure of the Representations of the $N=1$, $N=2$ Superconformal Algebras,''
  Int.\ J.\ Mod.\ Phys.\ A {\bf 3}, 1871 (1988).
  
  \bibitem{EOT}
 T.~Eguchi, H.~Ooguri and Y.~Tachikawa,
  ``Notes on the K3 Surface and the Mathieu group $M_{24}$,''
  Exper.\ Math.\  {\bf 20}, 91 (2011)
  [arXiv:1004.0956 [hep-th]].
  

  

\end{thebibliography}
\end{document}